\documentclass[aps,reprint, prl, superscriptaddress, longbibliography,showpacs,nobibnotes,floatfix]{revtex4-1}
\usepackage{graphicx}
\usepackage{bm,amsmath}
\usepackage{dcolumn}
\usepackage{natbib,xcolor}
\usepackage[version=4]{mhchem}
\usepackage{upgreek}
\usepackage{color}
\usepackage[colorlinks=true,linkcolor=blue,citecolor=blue,urlcolor=blue]{hyperref}
\usepackage{pifont}

\begin{document}
\title{Dominant in-plane anomalous Hall effect in a monoclinic room-temperature ferromagnet}

\author{Guoxin Zheng}
\affiliation{Division of Physics, Mathematics and Astronomy, California Institute of Technology, Pasadena, California 91125, USA}
\affiliation{Institute of Quantum Information and Matter, California Institute of Technology, Pasadena, CA 91125, USA}
\author{Arjyama Bordoloi}
\affiliation{Department of Mechanical Engineering, University of Rochester, Rochester, New York 14627, USA}
\author{Mingjun Fan}
\affiliation{Division of Physics, Mathematics and Astronomy, California Institute of Technology, Pasadena, California 91125, USA}
\author{Shunsuke Kitou}
\affiliation{Department of Advanced Materials Science, University of Tokyo, Kashiwa, 277-8561, Japan}
\author{Hiraku Saito}
\affiliation{Institute for Solid State Physics, University of Tokyo, Kashiwa, 277-8581, Japan}
\author{Taro Nakajima}
\affiliation{Institute for Solid State Physics, University of Tokyo, Kashiwa, 277-8581, Japan}
\affiliation{Institute of Materials Structure Science, High Energy Accelerator Research Organization, Tsukuba, Ibaraki 305-0801, Japan}
\affiliation{RIKEN Center for Emergent Matter Science, Wako 351-0198, Japan}
\author{Sobhit Singh}
\affiliation{Department of Mechanical Engineering, University of Rochester, Rochester, New York 14627, USA}
\affiliation{Materials Science Program, University of Rochester, Rochester, New York 14627, USA}
\author{Takashi Kurumaji}
\email{kurumaji@caltech.edu}
\affiliation{Division of Physics, Mathematics and Astronomy, California Institute of Technology, Pasadena, California 91125, USA}
\affiliation{Institute of Quantum Information and Matter, California Institute of Technology, Pasadena, CA 91125, USA}
\author{Linda Ye}
\email{lindaye@caltech.edu}
\affiliation{Division of Physics, Mathematics and Astronomy, California Institute of Technology, Pasadena, California 91125, USA}
\affiliation{Institute of Quantum Information and Matter, California Institute of Technology, Pasadena, CA 91125, USA}

\date{\today}
\maketitle
\textbf{Ferromagnetic metals are characterized by enhanced dissipationless transverse transport responses via the anomalous Hall effect, offering a route towards magnetic sensing and spintronic readout functionalities. In most ferromagnets, the anomalous Hall current is constrained to lie in the plane perpendicular to the magnetization (or applied magnetic field). Recently, it has been recognized that selected symmetries can also permit a Hall response in a traditionally forbidden configuration, where the Hall current lies in the same plane as the magnetization, realizing an in-plane anomalous Hall effect. Reported realizations of this effect, however, are typically much weaker than the conventional Hall response in the same material. Here, through engineering specific crystallographic mirror symmetry-breaking, we realize a strongly enhanced in-plane anomalous Hall response in monoclinic \ce{Cr3Te4} with room-temperature ferromagnetism. Remarkably, the in-plane anomalous Hall signal exceeds the out-of-plane response by a factor of five, with which we demonstrate a unique in-plane field and current sensing functionality. Combined with density functional theory calculations, our results establish low-crystalline-symmetry ferromagnets with near-Fermi-level Weyl points as a practical platform for symmetry-engineered Hall responses, and point to a route towards room-temperature, geometry-flexible sensing devices.}

\begin{figure*}[t]
	\includegraphics[width =  \textwidth]{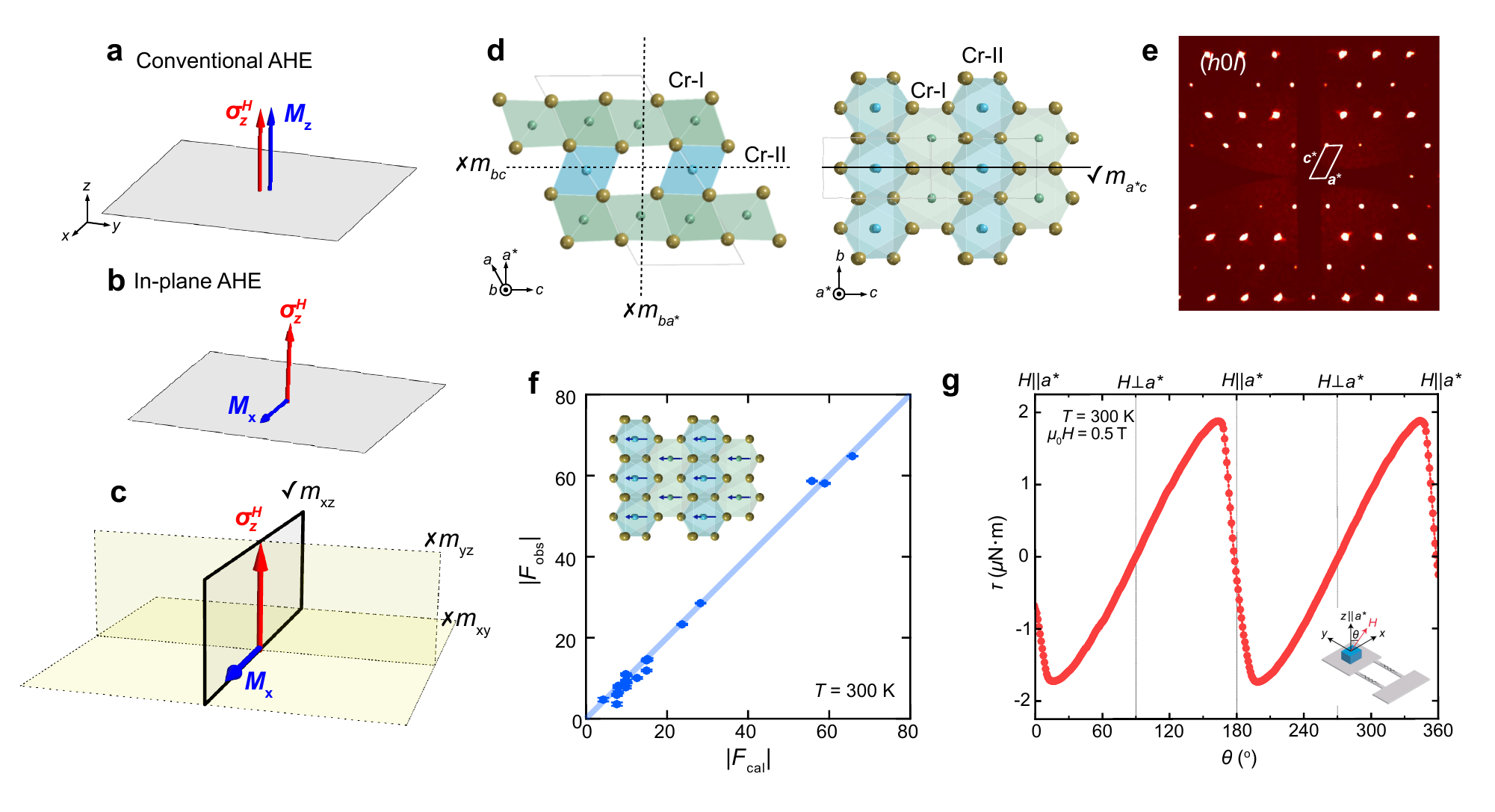}
	\caption{\label{fig1}
    \textbf{In-plane anomalous Hall effect, (broken) mirror symmetries, and monoclinic room-temperature ferromagnet \ce{Cr3Te4} }
    \textbf{a} Conventional out-of-plane anomalous Hall effect (OPAHE): an out-of-plane magnetization $M_z$ introduces a finite out-of-plane anomalous Hall conductivity $\sigma^{\rm H}_z$. Here $\sigma^{\rm H}_z$ corresponds to transverse current flow within the $xy$ plane.
    \textbf{b} In-plane anomalous Hall effect (IPAHE): a finite $\sigma^{\rm H}_z$  can be introduced by an in-plane magnetization $M_x$ when permitted by symmetry.
    \textbf{c} An example set of symmetries that allows IPAHE (see text): here crystallographic mirror symmetries $m_{xy}$ and $m_{yz}$ are broken, while $m_{xz}$ is preserved.
    \textbf{d} Crystal structure of monoclinic \ce{Cr3Te4}, where Te atoms are shown as brown spheres, and the two crystallographically distinct Cr sites (Cr$_{\rm I}$ and Cr$_{\rm II}$) are shown as green and blue spheres, respectively.
    Preserved $m_{a^*c}$ mirror plane is highlighted as solid line and broken mirror planes $m_{bc}$ and $m_{ba^*}$ are shown in dashed lines.
 \textbf{e} Reconstructed precession images of the reciprocal lattice space for $(h0l)$ plane.
 White rhombus indicates the reciprocal lattice vectors.
    \textbf{f} Observed $|F_{\text{obs}}|$ and calculated $|F_{\text{cal}}|$ structure factor amplitudes for \ce{Cr3Te4} at $T=300$ K with $\mu_0H=1$ T along the $c$ axis, obtained from neutron diffraction.
    $|F_{\text{obs}}|$ is scaled by a single parameter.
    $|F_{\text{cal}}|$ is obtained based on the FM model with moments along $c$ axis (inset, see SI Sec.~II for detail).
    The solid line indicates $|F_{\text{obs}}|=|F_{\text{cal}}|$.
    \textbf{g} Angular dependence of magnetic torque $\tau$ at $T=300$~K and $\mu_0H=0.5$~T, where $\theta$ is the angle between $a^*$ and $\bm{H}$, as shown in the inset.
 }
\end{figure*}

\begin{figure*}[t]
	\includegraphics[width =  0.9\textwidth]{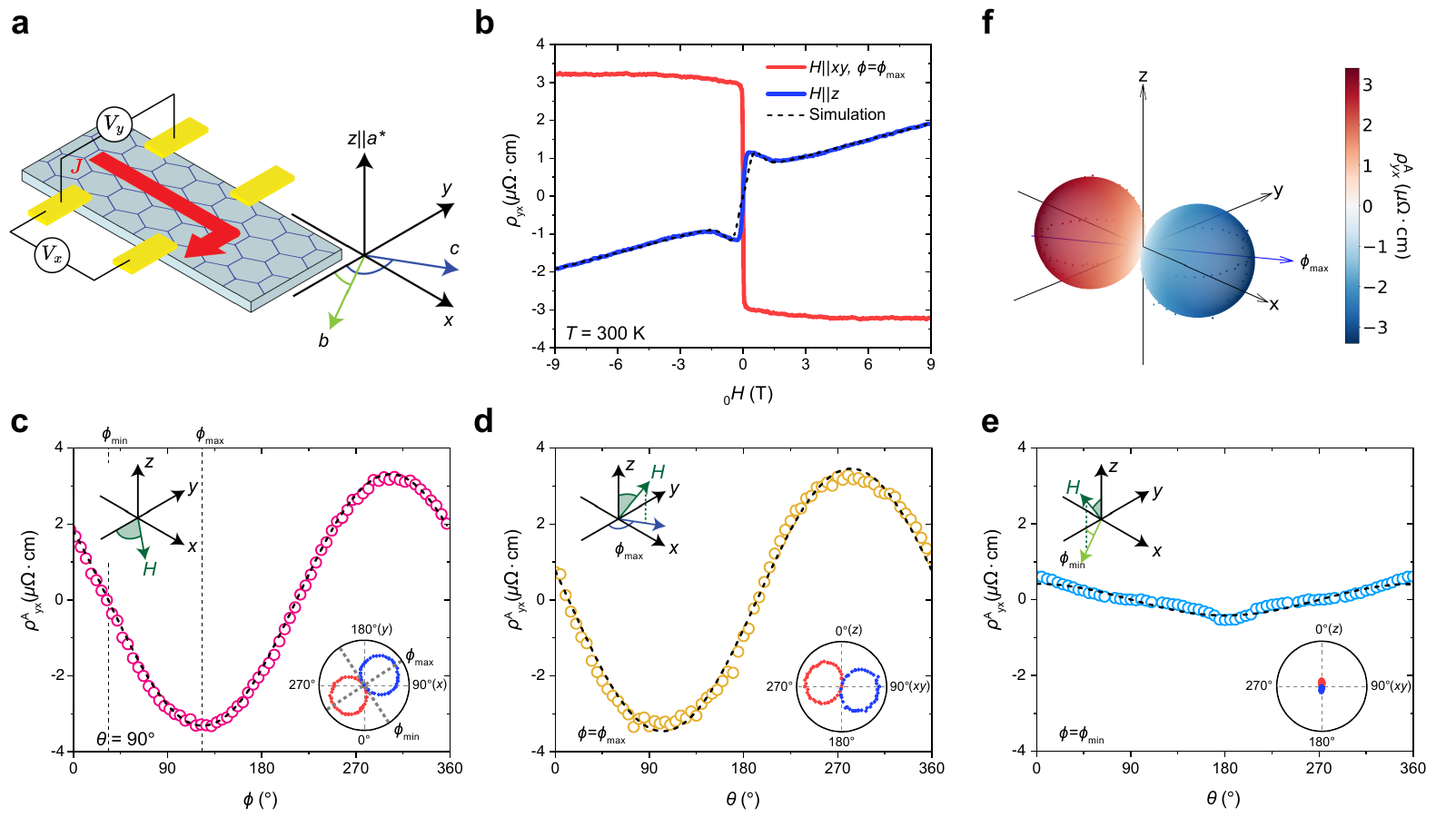}
	\caption{\label{fig2}
    \textbf{In-plane anomalous Hall effect in \ce{Cr3Te4} }
    \textbf{a} Schematic of the Hall measurement.
    The transport plane is the crystallographic $bc$ plane (defined as the $xy$ plane), with current $J\parallel x$, which is parallel to the crystallographic $b$ direction of one of the domains; the transverse voltage is captured along $y$ and the out-of-plane direction is $z\parallel a^*$.
    \textbf{b} The magnetic field dependence of Hall resistivity $\rho_{yx}$ at $T=300$~K for $\bm{H}\parallel z$ (blue curve) and $\bm{H}\parallel xy$ applied at $\phi=\phi_{\rm max}$ (red curve).
    The black dashed curve is a fit describing a crossover behavior from IPAHE at low field and OPAHE at high field (see text and SI).
    \textbf{c-e} Angular dependence of the anomalous Hall resistivity $\rho_{yx}^{\rm A}$ measured at $T=300$~K for magnetic field rotated in three mutually orthogonal planes (respective measurement geometries sketched as upper left insets), where $\theta$ is defined as the angle between $\bm{H}$ and $z$ and $\phi$ is defined as the angle between $\bm{H}$ and $-y$. Dashed lines are sinusoidal fits based on Eq. \ref{eq_rho_yx_ang} (see text).
   The lower right insets in \textbf{c-e} summarize the respective angular evolution of $\rho_{yx}^{\rm A}$ in polar diagrams.
    Red and blue colors indicate positive and negative $\rho_{yx}^{\rm A}$, respectively.
    \textbf{f} Three-dimensional angular map of $\rho_{yx}^{\rm A}$ constructed from the rotation scans in \textbf{c-e}, with the same color convention as in \textbf{c-e}.
 }
\end{figure*}

\begin{figure*}[t]
	\includegraphics[width =  0.9\textwidth]{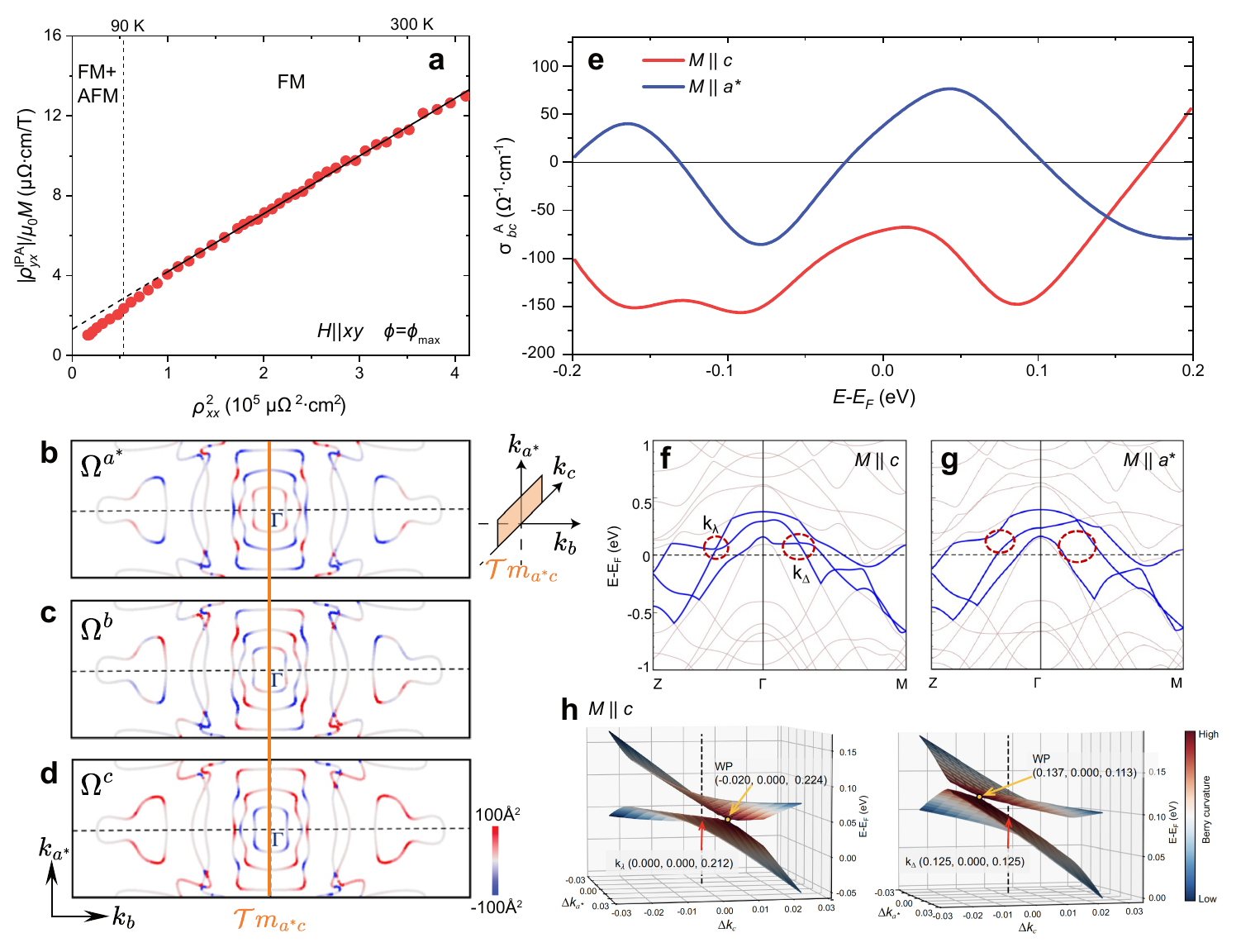}
	\caption{\label{fig3}
    \textbf{Berry curvature and Weyl points with in-plane magnetization in \ce{Cr3Te4}}
    \textbf{a}, $\rho_{xx}^2$-dependence of $\rho_{yx}^{\rm IPA}/(\mu_0 M)$. The dashed black line is the linear fit for data obtained between 140 K to 320 K. 
    \textbf{b-d}, Berry curvature-projected Fermi surfaces calculated for $\bm{M} \parallel c$ on the $k_{a^*}$--$k_b$ plane for $\Omega^{a^*}$, $\Omega^{b}$, and $\Omega^{c}$, respectively. The Berry curvature is plotted on a red-blue scale, with red (blue) denoting positive (negative) values. The orange line in \textbf{b-d} marks the magnetic symmetry $\mathcal{T}m_{a^* c}$.
    \textbf{e}, theoretically calculated in-plane anomalous Hall conductivity $\sigma^A_{bc}$ as a function of chemical potential near the Fermi level ($E_F$) for $\bm{M}\parallel c$ (red) and $\bm{M}\parallel a^*$ (blue).
    \textbf{f,g}, Electronic band structures calculated with spin-orbit coupling along selected high-symmetry $k$ directions for in-plane magnetization $\bm{M} \parallel c$ (\textbf{f}) and out-of-plane magnetization $\bm{M} \parallel a^*$ (\textbf{g}). Bands near the Fermi level  are highlighted in blue; red circles indicate the regions where valence and conduction bands touch each other and generate Weyl points (WP) near the Fermi level in the $\bm{M} \parallel c$ case. 
    \textbf{h}, Band dispersion near the WPs for $\bm{M}\parallel c$, computed on a dense 2D $k_{a^*}$--$k_c$ grid centered at $k_\lambda$ (0,\,0,\,0.212) and $k_\Delta$ (0.125,\,0,\,0.125) in fractional coordinates situated along $\Gamma \rightarrow Z$ and $\Gamma \rightarrow M$ high symmetry $k$-directions in the Brillouin zone, respectively.  The exact locations of the WPs are marked by yellow circles. Color represents the projection of normalized Berry curvature on the bands forming WPs. 
    }
\end{figure*}
The anomalous Hall effect (AHE), often defined as a transverse transport response generated by ferromagnetic orders, derives directly from time reversal symmetry-breaking and have profound connections with band topology and Berry curvature in the momentum space~\cite{nagaosa2010anomalous, xiao2010berry}. 
In most cases, AHE involves three mutually orthogonal vectors---magnetization ($\bm{M}$), electric current ($\bm{J}$), and transverse electric field ($\bm{E}$)~\cite{nagaosa2010anomalous}. Defining the corresponding Hall conductivity $\bm{\sigma}^{\rm H}$ as a vector that is perpendicular to both $\bm{J}$ and $\bm{E}$ \cite{liu2025multipolar}, the conventional AHE corresponds to $\bm{\sigma}^{\rm H}$ orienting parallel with $\bm{M}$ (Fig. \ref{fig1}a).
In recent years, there is a growing understanding that an in-plane $\bm{M}$ can also generate a perpendicular $\bm{\sigma}^{\rm H}$ (Fig.~\ref{fig1}b), realizing an in-plane anomalous Hall effect (IPAHE)  \cite{zhou2022heterodimensional,chen2022unconventional,wang2024orbital,nakamura2024plane,wang2025parallel, kao2025magnetization,sankar2025room}. 

To arrive at this unconventional Hall configuration, here we focus on
crystallographic and magnetic symmetries, which have been discussed to play central roles in constraining the relative orientations of the anomalous Hall conductivity $\bm{\sigma}^{\rm H}$ and magnetization $\bm{M}$~\cite{tan2021unconventional,kurumaji2023symmetry,cao2023plane,liu2025multipolar}.
Taking the crystallographic mirror symmetry as an example, the presence of a single mirror plane constrains the two vectors to be simultaneously either even or odd with respect to the mirror operation; as both $\bm{\sigma}^{\rm H}$ and $\bm{M}$ are pseudo axial vectors, they are required to be both perpendicular to, or both lying within, the mirror plane. In the presence of two or more mutually perpendicular mirror planes,  $\bm{\sigma}^{\rm H}$ will be dictated to be parallel with $\bm{M}$ as in the conventional anomalous Hall effect \cite{tan2021unconventional,kurumaji2023symmetry}.
To date IPAHE has been reported in a number of low symmetry systems~\cite{zhou2022heterodimensional,wang2025parallel,kao2025magnetization,sankar2025room,wang2024orbital,nakamura2024plane}; in most cases, however, the IPAHE is substantially weaker than the corresponding conventional anomalous Hall response, underscoring the need for a material platform in which a robust in-plane Hall response dominates over the conventional Hall effect.

In this work, we combine targeted crystallographic symmetry breaking with room-temperature ferromagnetic order to realize a large in-plane anomalous Hall effect that dominates over the conventional out-of-plane anomalous Hall response in the layered monoclinic crystal \ce{Cr3Te4}.
The specific symmetry configuration employed to generate an IPAHE with dominant out-of-plane Hall $\sigma^{\rm H}_z$ and in-plane magnetization $M_x$ is illustrated in Fig.~\ref{fig1}c.
In this configuration, the preserved mirror symmetry $m_{xz}$ permits the coexistence of $\sigma^{\rm H}_z$ and $M_x$ (both odd under $m_{xz}$), while the absence of $m_{xy}$ and $m_{yz}$ lifts any further incompatibility between $\sigma^{\rm H}_z$ and $M_x$. We emphasize that the mirror symmetries considered here are purely crystallographic and should be distinguished from mirror operations in magnetic space or point groups; this framework enables us to analyze the symmetry constraints on the Hall tensor imposed by the crystal structure without \textit{a priori} assumptions about the magnetic order.
The particular combination of preserved and broken mirror symmetries is naturally realized in monoclinic \ce{Cr3Te4}, which crystallizes in space group $C2/m$  (Fig.~\ref{fig1}d, see SI Sec.~I).
The crystal structure is composed of 1T-\ce{CrTe2} layers of edge-sharing \ce{CrTe6} octahedrons (we label these Cr atoms as Cr-I), while in between the layers the gap is intercalated with one-dimensional chains of Cr atoms (Cr-II) along the $b$ direction~\cite{wang2023field}; the structure lacks mirror symmetries $m_{bc}$ and $m_{ba^*}$, 
while $m_{a^*c}$ is preserved. 
The monoclinic unit cell is also evident in single crystal X-ray diffraction pattern shown in Fig.~\ref{fig1}e. Viewed in conjunction with the schematic in Fig.~\ref{fig1}c, for electrical transport within the $bc$ plane, a magnetization oriented along the $c$ axis in principle permits a finite $\sigma^{\rm H}_z$, thereby enabling an in-plane Hall response.

As-grown crystals of \ce{Cr3Te4} are of plate-like shape with large $bc$ planes (see SI Sec.~I). 
The magnetic properties of our \ce{Cr3Te4} crystals are found to be consistent with previous reports~\cite{wang2023field,bose2025anomalous,bhartiya2025competing,purwar2023investigation}, exhibiting a paramagnetic to ferromagnetic (FM) phase transition at $T_c\sim322$~K and a coexistence of ferromagnetic and antiferromagnetic (AFM) phases below $\sim90$~K (see SI Sec.~IV).
Magnetization measurements further indicate an anisotropic ferromagnetic state, with $bc$ as the easy plane and $a^*$ as the hard axis (see SI Sec. IV). The easy-plane nature of the ferromagnetic order at room temperature is verified by neutron scattering (Fig.~\ref{fig1}f), where the refined structure points to ferromagnetic moments on both Cr-I and Cr-II sites aligned along the crystallographic $c$ axis (see SI Sec.~II for details), consistent with an early neutron study \cite{andresen1970magnetic}. Fig.~\ref{fig1}g shows angular dependence of magnetic torque of a \ce{Cr3Te4} crystal, where the saw-tooth pattern and the sharp switching behavior near $\bm{H}\parallel a^*$ further verifies the easy-plane anisotropy perpendicular to $a^*$~\cite{modic2017robust}.\\

\noindent\textbf{Observation of IPAHE at room temperature} Here we present the central result of this work: the observation of IPAHE in \ce{Cr3Te4} at room temperature.
We note that the hexagonal-shaped crystals of \ce{Cr3Te4} used in our study are composed of twinning of three inequivalent monoclinic domains that share their $a^*$ axis and have their principal $b$ axes (Cr-II chain direction) approximately $120^{\circ}$ from each other (see SI Sec.~I). We will return to the effects of twinning on the IPAHE below.
The configuration for the transport measurements is shown in Fig.~\ref{fig2}a (also see Methods). For all experiments shown below, the currents are confined within the crystallographic $bc$ plane, i.e., the \ce{CrTe2} layers.
We define the current $\bm{J}$ direction as the $x$ axis, which is parallel to the crystallographic $b$ axis of one of the domains. The Hall voltage is measured along $y$ and the out-of-plane direction is defined as the $z$ axis with $z\parallel a^*$.

Figure~\ref{fig2}b shows representative Hall resistivity $\rho_{yx}$ measured as a function of applied magnetic field $H$ oriented within the $xy$ plane (red curve) and perpendicular to the $xy$ plane (blue curve) at $T = 300$ K.
For the selected orientation of in-plane field, $\rho_{yx}$ exhibits a prototypical anomalous Hall response that closely tracks the in-plane magnetization (see SI Sec. IV), with a negligible linear-in-$H$ ordinary Hall component.
In contrast, for out-of-plane fields, $\rho_{yx}$ is substantially smaller in magnitude and displays a non-monotonic cusp-like feature.
Such behavior was previously reported and attributed to a topological Hall effect \cite{purwar2023investigation}.
Here, however, in light of the easy-plane anisotropy and the dominant in-plane anomalous Hall effect, we find that the field evolution of $\rho_{yx}$ can be quantitatively described by an initial sharp increase arising from in-plane magnetic domain alignment and the IPAHE, followed by a gradual rotation of the magnetic moments toward the applied field and the hard $a^*$ axis (black dashed line in Fig.~\ref{fig2}b; see SI Sec.~V).
An ordinary Hall component is also resolved by the $H$-linear slope at $H>2$ T in the out-of-plane configuration.
Within this framework, we extract the out-of-plane anomalous Hall resistivity $\rho_{yx}^{\mathrm{OPA}} = 0.6~\mu\Omega\cdot\mathrm{cm}$ and in-plane anomalous Hall resistivity $\rho_{yx}^{\mathrm{IPA}} = 3.2~\mu\Omega\cdot\mathrm{cm}$ at the high field limit, where the IPAHE exceeds the conventional anomalous Hall effect by nearly a factor of five.

Next we turn to evaluating the angular dependence of the anomalous Hall responses in \ce{Cr3Te4} with rotation of $H$.
For clarity, we focus on the anomalous component $\rho_{yx}^{\rm A}$ extracted from high-field regime above the respective saturation field at all orientations. The full angular evolution of $\rho_{yx}^{\rm A}$ is summarized in Figs.~\ref{fig2}c–e (see raw data in SI Sec. VIII). Here $\theta$ is the angle between $\bm{H}$ and $+z$, $\phi$ is the azimuthal angle of $\bm{H}$ measured from $-y$ within the $bc(xy)$ plane. Fig. \ref{fig2}c highlights results obtained from rotation of magnetic field within the $xy$ plane, where $\rho_{yx}^{\rm A}$ exhibits a sinusoidal behavior which maximizes near $\phi_{\text{max}}\sim123.3^\circ$, pointing to that the IPAHE is dominated by the component of in-plane magnetic moment along $\phi_{\text{max}}$.
Figures \ref{fig2}d and \ref{fig2}e show polar-angle dependence measured near $\phi_{\text{max}}$ and $\phi_{\text{min}}$, respectively. We find that the full $(\theta,\phi)$ dependence of the Hall resistivity is well described by the expression (dashed fits in Figs. \ref{fig2}c–e)
\begin{equation}
\rho_{yx}^{\rm A} = \rho_{yx}^{\rm OPA}\cos\theta + \rho_{yx}^{\rm IPA}\sin\theta \cos(\phi - \phi_{\rm max}),
\label{eq_rho_yx_ang}
\end{equation}
which reflects a decomposition of the anomalous Hall response into conventional out-of-plane and the new in-plane components that couple to the $z$-component of the magnetization and to the in-plane magnetization component along $\phi_{\text{max}}$, respectively.
The complete three-dimensional angular dependence of $\rho_{yx}^{\rm A}$ via a spherical-harmonic fit is summarized in Fig.~\ref{fig2}f, again highlighting that the maximum anomalous Hall response occurs near the $xy$ plane. 
We further note that the extracted phase offset $\phi_{\rm max}\approx 120^\circ$ indicates that one of the three $120^{\circ}$-domains is dominant for the observed IPAHE (see SI Sec.~VI) and $\phi_{\rm max}$ is near the $c$-axis of the `primary' domain (Fig.~\ref{fig1}d). 
In this context, the overall response (Fig.~\ref{fig2}f and Eq. \ref{eq_rho_yx_ang}) is fully compatible with the monoclinic symmetry of \ce{Cr3Te4} (Fig. \ref{fig1}d): with $\phi_{\rm min}$ lying near the $b$ axis of the principal domain, $\bm{M}$ becomes incompatible with $\sigma_z^{\rm H}$ with respect to $m_{a^*c}$ and thereby yielding a vanishing Hall response. The IPAHE in \ce{Cr3Te4} has been confirmed in two additional samples, with consistent behaviour (see SI Sec. VII).\\

\begin{figure*}
    \centering
    \includegraphics[width=0.8\linewidth]{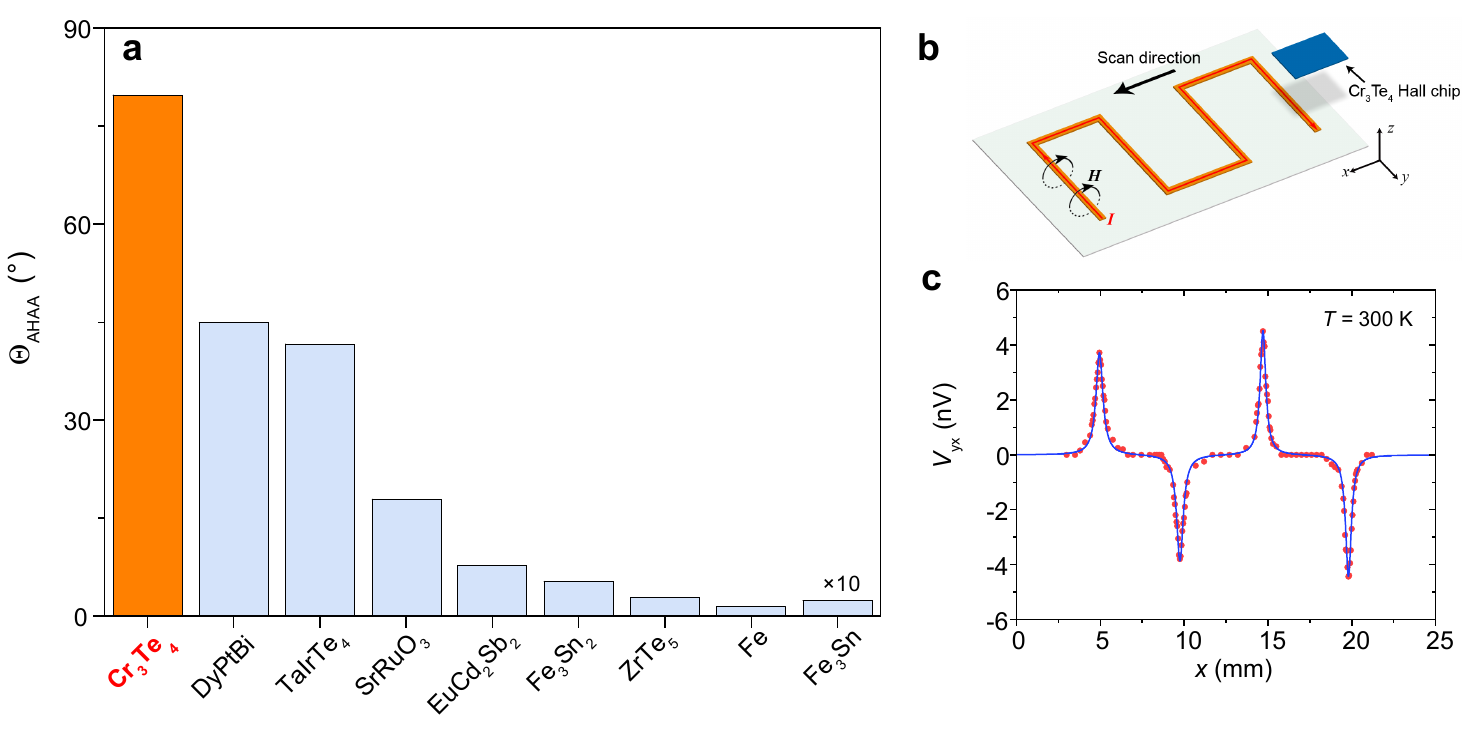}
    \caption{\textbf{Anomalous Hall anisotropy and current sensing enabled by IPAHE in \ce{Cr3Te4}}
    \textbf{a}, Comparison of the anomalous Hall anisotropy angle, $\Theta_{\rm AHAA}\equiv\tan^{-1}(\sigma^{\rm IPA}/\sigma^{\rm OPA})$, of \ce{Cr3Te4} (this work, orange bar) with representative systems previously reported to exhibit in-plane anomalous Hall effects, including DyPtBi \cite{chen2022unconventional}, \ce{TaIrTe4}/CGT heterostructures \cite{kao2025magnetization}, \ce{SrRuO3} \cite{nishihaya2025spontaneous}, \ce{EuCd2Sb2} \cite{nakamura2024plane}, \ce{Fe3Sn2} \cite{wang2024orbital}, \ce{ZrTe5} \cite{wang2025parallel} ,  Fe \cite{peng2024observation}, and \ce{Fe3Sn} \cite{sankar2025room} (blue bars).  $\Theta_{\rm AHAA}$ of \ce{Fe3Sn} has been enlarged by a factor of 10 for clarity. 
    \textbf{b}, Schematic of the experiment utilizing \ce{Cr3Te4} as an in-plane Hall sensor. A \ce{Cr3Te4} single-crystal is placed parallel to and above a patterned current-carrying circuit and scanned along $x$ direction.
    \textbf{c}, Measured Hall voltage $V_{yx}$ during a representative scan, showing sharp peaks as the sensor crosses the current-carrying wires. The blue curve represents calculation based on the magnetic field generated by the known current distribution (see Methods).
    }
    \label{fig4}
\end{figure*}

\noindent\textbf{Intrinsic origin of IPAHE in \ce{Cr3Te4}} 
The conventional AHE is well understood to be broadly categorized into intrinsic and extrinsic contributions~\cite{nagaosa2010anomalous,tian2009proper}, which can be empirically analyzed using the Tian-Ye-Jin model~\cite{tian2009proper,hou2015multivariable}. Here we adpot a similar framework to obtain insights into our observed IPAHE. With the strong temperature evolution of magnetization in \ce{Cr3Te4} up to room temperature, 
we normalize  $\rho_{yx}^{\rm IPA}$ by $\mu_0 M$ to compare the temperature evolution of  $\rho_{yx}^{\rm IPA}$ with  $\rho_{xx}$:
\begin{align}
\rho_{yx}^{\rm IPA}/(\mu_0 M)=a^\prime\rho_{xx0}+b^\prime\rho_{xx}^2,
\label{eq_scaling}
\end{align}
here $\rho_{xx0}$ is the residual longitudinal resistivity, $a^\prime$ is the skew-scattering coefficient associated with impurity scattering, and $b^\prime$ is an effective coefficient that captures the combined intrinsic contribution related to the Berry curvature and the side-jump contribution. 

$\rho_{yx}^{\rm IPA}/(\mu_0 M)$ with respect to $\rho_{xx}$ is shown in Fig.~\ref{fig3}a (raw data see SI Sec. IX).
The data points within the ferromagnetic phase between 140 and 320~K can be fitted by Eq.~\ref{eq_scaling} (solid black line), yielding $b^\prime=28.9~\Omega^{-1}\cdot\mathrm{cm}^{-1}\cdot\mathrm{T}^{-1}$.
On the basis of the assumption that the side-jump effect is negligible in the intrinsic regime, we obtain an estimated intrinsic in-plane anomalous Hall conductivity to be $8.2~\Omega^{-1}\cdot\mathrm{cm}^{-1}$ at $T=300$~K.
We note that the data deviate from the linear fit upon approaching $90$~K, suggesting additional contributions from the mixed ferromagnetic/antiferromagnetic phase.

To elucidate the origin of the in-plane AHE, we performed first-principles density functional theory (DFT) calculations to obtain the Berry curvature $\bm{\Omega}$ of the system. We first focus on the case with $\bm{M}\parallel c$, corresponding to where the experimental IPAHE is found to be maximized. 
Fig. \ref{fig3}b–d shows the Fermi surface projections of the Berry curvature components $\Omega^{a^*}$, $\Omega^b$, and $\Omega^c$, respectively. The full magnetic symmetry in this case contains an $\mathcal{T}m_{a^*c}$ operation ($\mathcal{T}$: time-reversal), which we highlight as an orange line in the momentum space in Fig.~\ref{fig3}b-d. Under $\mathcal{T}m_{a^*c}$,  $\Omega^b$ (Fig.~\ref{fig3}c) is dictated to be an odd function of $k_b$ and therefore vanishes upon integration over momentum space. By contrast, $\Omega^c$ (Fig. \ref{fig3}d), whose direction follows the magnetization axis $\bm{M}\parallel c$, is symmetry-allowed. Most importantly, $\Omega^{a^*}$ (Fig. \ref{fig3}b), which corresponds to the experimentally captured IPAHE in the $xy$ plane, is also symmetry allowed despite being transverse to the magnetization direction, and can be attributed to the imbalance between the red (positive) and blue (negative) Berry-curvature hot spots.

Motivated by the dominant IPAHE over the corresponding OPAHE, we further examine the dependence of the Hall conductivity with the orientation of ferromagnetic moments.
Figure~\ref{fig3}e shows the DFT-calculated anomalous Hall conductivity within the $bc$ plane $\sigma_{bc}^{\rm A}$ (corresponding to the experimental configuration $\sigma_{xy}$) as a function of energy near the Fermi level $E_F$ for $\bm{M}\parallel c$ (red) and $\bm{M}\parallel a^*$ (blue).
$\sigma_{bc}^{\rm A}$ is finite for both $\bm{M}\parallel c$ and $\bm{M}\parallel a^*$, with the former being substantially larger than the latter for a relatively broad window near the Fermi level. 

To elucidate the origin of the enhanced anomalous Hall conductivity with in-plane magnetization, we compare the electronic band structures for $\bm{M}\parallel c$ and $\bm{M}\parallel a^*$ in Fig.~\ref{fig3}f and g. 
Pronounced magnetic moment orientation-dependence can be found in the band dispersion along the high-symmetry $k$-paths Z\,$\rightarrow$\,$\Gamma$$\rightarrow$\,M (see SI Sec.~III for the full band structures and Brillouin zone).
Near the Fermi level, for $\bm{M}\parallel c$, the bands highlighted in blue approach each other closely in the regions marked by dashed circles, forming near-touching points indicative of gapless Weyl points (WPs) in their vicinity, whose exact locations are shown in Fig.~\ref{fig3}h.
These WPs act as singular sources of Berry curvature and thereby leading to the enhanced anomalous transport. In contrast, for $\bm{M}\parallel a^*$ (Fig.~\ref{fig3}g), the same bands remain well separated and do not appear to intersect. We therefore attribute the enhanced IPAHE over OPAHE in \ce{Cr3Te4} to the presence of gapless WPs right near the Fermi level for $\bm{M}\parallel c$. \\

\noindent\textbf{Anomalous Hall anisotropy and in-plane field detection}
To put the IPAHE in \ce{Cr3Te4} in broader context of systems where in-plane Hall effects have been reported, we define an anomalous Hall anisotropy angle $\Theta_{\rm AHAA}=\tan^{-1}(\sigma^{\rm IPA}/\sigma^{\rm OPA})$ to quantify the relative strength of IPAHE and OPAHE.
This yields $\Theta_{\rm AHAA}=79.4^\circ$ for \ce{Cr3Te4}, which is the largest compared with previously reported cases, e.g., \(\Theta_{\rm AHAA}\sim 5^\circ\) in the kagome
ferromagnet \ce{Fe3Sn2}, \(\Theta_{\rm AHAA}\sim 7.8^\circ\) in the zintl antiferromagnet \ce{EuCd2Sb2}, \(\Theta_{\rm AHAA}\sim41^\circ\) in \ce{TaIrTe4}/\ce{Cr2Ge2Te6} heterostructures, and \(\Theta_{\rm AHAA}\sim17.9^\circ\) in \ce{SrRuO3} thin films \cite{chen2022unconventional,wang2024orbital,nakamura2024plane,wang2025parallel, kao2025magnetization,sankar2025room, peng2024observation,nishihaya2025spontaneous} (Fig. \ref{fig4}a). 

The uniquely large $\Theta_{\rm AHAA}$ in \ce{Cr3Te4} indicates its utility as a novel in-plane Hall sensor at low magnetic fields.
Figure~\ref{fig4}b schematically shows a setup for a proof-of-principle demonstration of such capability. A piece of single-crystalline \ce{Cr3Te4}, serving as the in-plane Hall sensor, is placed parallel to and above a circuit board carrying a patterned current. As discussed above, the sensor selectively detects in-plane magnetic fields along $\phi_{\rm max}$. 
As the device is scanned across the current-carrying wires, the in-plane component of the Oersted field generated by the wires induces a measurable Hall voltage owing to the large $\Theta_{\rm AHAA}$.  Figure~\ref{fig4}c shows the Hall voltage of our \ce{Cr3Te4} device during a single $x$-scan, revealing sharp peaks as the sensor travels above the current-carrying wires. The blue curve is the calculated result based on the known current distribution (see Methods). Notably, the test is performed entirely at room temperature. The sensitivity of such in-plane Hall sensing devices can be further improved when fabricated into thin film geometries. Together, these results highlight large $\Theta_{\rm AHAA}$-materials like \ce{Cr3Te4} as promising platforms for geometry-flexible magnetic sensing. \\

\noindent\textbf{Discussion and Conclusion}

In summary, through engineering the interplay between crystalline symmetry and the ferromagnetic order in monoclinic \ce{Cr3Te4}, we have established a material platform where the in-plane anomalous Hall responses dominates over the conventional, out-of-plane counterpart.
Conceptually, our case study demonstrates how real-space symmetry engineering can be used to tailor the momentum-space Berry curvature distribution, producing a strongly anisotropic Hall response.
Notably, when viewed together with the large linear in-plane Hall effect reported in the VS$_2$–VS superlattice, which exhibits a Hall anisotropy angle of $51.3^\circ$ \cite{zhou2022heterodimensional}, monoclinic symmetry appears particularly favorable for promoting strong in-plane Hall responses. This contrasts with rhombohedral systems, where multiple non-orthogonal mirror planes tend to constrain the form of the Hall tensor and suppress such anisotropy \cite{wang2024orbital,nakamura2024plane}. The intrinsic bulk symmetry breaking in \ce{Cr3Te4} may also play an important role in enhancing the in-plane response, compared with systems in which mirror symmetry is broken only at surfaces or interfaces \cite{sankar2025room,peng2024observation,nishihaya2025spontaneous}.

More broadly, the symmetry-based design principles established here may provide a route toward engineering quantized anomalous Hall states controlled by in-plane ferromagnetic order \cite{liu2013plane,Liu2018-dx}, Hall responses arising from interplay between antiferromagnetic orders and lattice symmetries \cite{Smejkal2020-so}, and magnetic field-induced quantum geometric effects associated with anomalous orbital polarizability \cite{Wang2024-vc,wang2024orbital}. Overall, our results highlight symmetry-engineered magnetic materials as a promising platform for tunable Berry curvature physics and next-generation Hall effect-based electronic functionalities.
 
\section{Methods} 
\subsection{Crystal synthesis} 
Single crystals of \ce{Cr3Te4} were grown by chemical vapor transport reaction with \ce{I2} as the transport agent~\cite{yamaguchi1972magnetic,goswami2024critical}.
Stoichiometric amount of chromium chunk (99.99\%) and tellurium shot (99.99\%) were used as received without further purification.
Two gram of the above mixture together with approximately 50 mg of \ce{I2} were loaded into an evacuated quartz tube.
The quartz tube was placed in a three-zone furnace with the temperature gradient between 1000$^{\circ}$C (charged zone) and 820$^{\circ}$C (crystallization zone).
This temperature gradient was held for seven days.
Obtained single crystals were black shiny and hexagonal plate shape, which are typically 1-3 mm in dimensions. 

\subsection{Structural characterization}
The atomic composition was checked by energy dispersive x-ray spectroscopy (EDX) using ZEISS 1550VP field emission SEM with Oxford X-Max SDD X-ray Energy Dispersive Spectrometer.
Single-crystal X-ray diffraction data was collected on Rigaku XtaLAB diffractometer equipped with a HyPix-6000HE detector using Mo $K_{\alpha}$ radiation at room temperature.
Details of the structural analysis results are shown in supplementary information (SI Sec.~I).

\subsection{Physical property measurements}
The magnetization ($M$) was measured with a Vibrating Sample Magnetometer and Heat Capacity option implemented on a commercial superconducting magnet (Quantum Design DynaCool Physical Properties Measurement System, PPMS).
The magnetic torque ($\tau$) was measured with the Torque Magnetometer option. $\tau$ is given by $\boldsymbol{\tau}=\mu_0V_0\,\bm{M}\times\bm{H}$, where $V_0$ is the sample volume.
Electrical transport measurements were performed with a conventional five-probe method with AC lock-in amplifiers at a typical frequency near 37 Hz and an excitation current of 4 mA.
The obtained longitudinal and transverse resistivities were field-symmetrized and antisymmetrized, respectively, to correct contact misalignment.

The anomalous Hall resistivity $\rho_{yx}^{\rm A}$ in Figs.~\ref{fig2}c-e was obtained by subtracting the linear ordinary Hall background from the measured $\rho_{yx}$ and taking the zero-field intercept of the resulting curve; the ordinary Hall background was determined from a linear fit to $\rho_{yx}$ in the range 6--9~T.

\subsection{Neutron diffraction}
Single-crystal neutron diffraction was carried out at a triple-axis spectrometer PONTA(5G) installed in JRR-3 of the Japan Atomic Energy Agency \cite{nakajima2024polarized}.
An incident neutron beam with the wavelength of 1.55 \AA $ $ was obtained by a pyrolytic graphite (PG) (002) monochrometer.
The spectrometer was operated in the two-axis mode, and the horizontal collimation was open-$80'-80'$.
An external magnetic field of 1.0 T was applied perpendicular to the scattering plane by a Helmholtz coil.

\subsection{Current sensing measurements}
The circuit board in Fig.~\ref{fig4}b was assembled by placing insulated copper wires with a diameter of 0.2 mm on a printed circuit board, and a DC current of $\pm 1$~A was supplied by a Keithley 2450 sourcemeter. The \ce{Cr3Te4} Hall chip was mounted on an $xyz$ translation stage, allowing controlled motion along three orthogonal directions. The Hall voltage was measured using a Stanford Research SR860 lock-in amplifier. Figure~\ref{fig4}c shows the antisymmetrized Hall voltage $V_{yx}$ with respect to the magnetic field. For a single straight wire carrying current $I$ along the $+y$ direction and located at $(x_0,-z_0)$, the $x$-component of the magnetic field is approximated as
\begin{align}
\mu_0H_x=\frac{\mu_0 I}{2\pi}\frac{z_0}{(x-x_0)^2+z_0^2}.
\label{eq_current_B}
\end{align}
In the low-field regime ($<10$~mT), the OPAHE is about 16 times smaller than the IPAHE (Fig.~\ref{fig2}b). The contribution from the out-of-plane field can therefore be neglected, such that $V_{yx}\approx k_A \mu_0 H_x$. Here, $k_A$ denotes the slope of the Hall voltage with respect to magnetic field, and is experimentally determined to be $19~\mu\mathrm{V}/\mathrm{T}$ from the field dependence of $V_{yx}$ measured in the PPMS. The spatial variation of the IPAHE signal during a scan along $x$ is therefore given by
\begin{align}
V_{yx}(x)\approx k_A\frac{\mu_0 I}{2\pi}\frac{z_0}{(x-x_0)^2+z_0^2}.
\label{eq_current_fit}
\end{align}
The blue curve in Fig.~\ref{fig4}c represents the calculated result based on Eq.~\ref{eq_current_fit}, using the measured value $k_A=19~\mu\mathrm{V}/\mathrm{T}$ and $z_0=0.25$, 0.24, 0.21, and 0.21 mm for current wires located at $x_0=5$, 9.8, 14.7, and 19.8 mm, respectively.

\section{Author Contribution}
G.Z. performed transport and magnetization measurements with T.K., and performed torque measurements.
The in-plane Hall sensing test was done by G.Z. and M.F.
A.B. and S.S. performed DFT calculations.
S.K. performed single crystal XRD.
T.K. synthesized single crystals, and performed neutron diffraction study with T.N. and H.S.
L.Y. and T.K. coordinated the project.

\section{Acknowledgments}
We thank Y.-T.Shao, M.T.Pamuk and K.Yi for fruitful discussions.
This study was partly supported by JST FOREST (No. JPMJFR2362), Ministry of Education Culture Sports Science and Technology (MEXT) Leading Initiative for Excellent Young Researchers (JPMXS0320200135) and Inamori Foundation. 
The neutron diffraction experiment was performed under proposal No. 25804.
The experimental works carried out at Caltech are supported by Gordon and Betty Moore Foundation through Moore Materials Synthesis Fellowship to L.Y. (GBMF12765) and Institute of Quantum Information and Matter (IQIM), an NSF Physics Frontier Center (PHY-2317110).
Part of the work is carried out at GPS Division Analytical Facility supported by NSF DMR-0080065, at the shared Physical Properties Measurement System facilities at Caltech supported by NSF DMR-2117094.
G.Z. acknowledges support from the Clark B. Millikan Postdoctoral Fellowship at Caltech.
S.K. thanks J. Yamaura for supporting the in-house X-ray diffraction measurements.
T.K. and T.N. thank H. Saito for supporting neutron experiment.
A.B.~and S.S.~acknowledge support from the U.S.~Department of Energy, Office of Science, Office of Fusion Energy Sciences, Quantum Information Science program under Award No.~DE-SC-0020340.
A.B.~and S.S.~thank support from the Furth Research Fund at the University of Rochester.
\bibliography{reference}

\end{document}